\documentclass[namedreferences]{solarphysics}

\usepackage[hyperref,optionalrh,showbiblabels]{spr-sola-addons} 
\usepackage{graphicx}        
\usepackage{color}           
\usepackage{breakurl}        

\newcommand{\aap}{    {\it Astron. Astrophys.}}

\newcommand{\apj}{    {\it Astrophys. J.}}

\newcommand{\pasj}{   {\it Pub. Astron. Soc. Japan}}

\newcommand{\solphys}{{\it Solar Phys.}}

\chardef\us=`\_

\begin{document}

\begin{article}
\begin{opening}

\title{Distributed Electric Currents in Solar Active Regions}

\author[email={yuriy.fursiak@gmail.com}]{\inits{Yu.A.}\fnm{Yuriy A.}~\lnm{Fursyak}}
\author[corref, email={alex.s.kutsenko@gmail.com}]{\inits{A.S}\fnm{Alexander S.}~\lnm{Kutsenko}}
\author[email={vabramenko@gmail.com}]{\inits{V.I}\fnm{Valentina I.}~\lnm{Abramenko}}

\address{Crimean Astrophysical Observatory, p/o Nauchny, Crimea, 298409, Russia}

\runningauthor{Yu.A.~Fursyak \textit{et al.}}
\runningtitle{Distributed Electric Currents in Solar ARs}

\begin{abstract}
Using magnetographic data provided by the \textit{Helioseismic and Magnetic Imager} on board the \textit{Solar Dynamics Observatory}, we analyzed the structure of magnetic fields and vertical electric currents in six active regions (ARs) with different level of flare activity.
We found that electric currents are well balanced over the entire AR: for all of them the current imbalance is below 0.1\%, which means that any current system is closed within an AR. 
Decomposition of the transverse magnetic field vector into two components allowed us to reveal the existence of large-scale vortex structures of the azimuthal magnetic field component around main sunspots of ARs. In each AR, we found a large-scale electric current system occupying a vast area of an AR, which we call distributed electric current.
For ARs obeying the Hale polarity law and the hemispheric helicity sign rule, the distributed current is directed upward in the leading part of an AR and it appears to be closing back to the photosphere in the following part of an AR through the corona and chromosphere. 
Our analysis of the time variations of the magnitude of the distributed electric currents showed that low-flaring ARs exhibit small variations of the distributed currents in the range of $\pm 20 \times 10^{12}$ A, whereas the highly flaring ARs exhibited significant slow variations of the distributed currents in the range of $30-95 \times 10^{12}$ A. Intervals of the enhanced flaring appear to be co-temporal with smooth enhancements of the distributed electric current.

\end{abstract}
\end{opening}

\section{Introduction}
     \label{S-Introduction}

It is widely accepted that the energy released during solar flares and coronal mass ejections is stored in active regions (ARs) in the solar corona in the form of the so-called ``free'' magnetic energy associated with the presence of electric currents \citep[\textit{e.g.}, ][to mention a few]{Abramenko1991, Melrose1991, Wang1996, Schrijver2005, Aschwanden2013, Fleishman2018,Toriumi2019}.
	
The issues that are actively discussed over the last decades are (i) the origin of these electric currents and (ii) whether these electric currents are neutralized. The neutralization of electric currents implies that no net current over one magnetic polarity of an AR at the photosphere level is present \citep{Wheatland2000}. In this case, the current system associated with a magnetic tube should consist of a direct current (presumably flowing in the central part of the tube) and a return (surface) current.
	
There are two ways in which electric currents may be built up in the corona. The first one is due to twisting or shearing of a magnetic flux tube by (sub)photospheric plasma motions \citep[\textit{e.g.}, ][]{McClymont1989, Torok2003, Aulanier2005, Dalmasse2015}. Alternatively, a current-carrying magnetic flux tube may emerge from beneath the photosphere \citep{Leka1996, Longcope2000}, \textit{i.e}, it could be twisted during its formation and/or during buoyant rising through the convection zone \citep[\textit{e.g.}, ][]{Cheung2014}.

Observations show that although electric currents integrated over the area of the entire AR are balanced to a good degree \cite[\textit{e.g.}, ][]{Abramenko1996,Schrijver2008,Georgoulis2012}, the current patterns remain non-neutralized at each magnetic polarity \citep{Georgoulis2012}. 
One evidence of a non-neutralized current pattern is the well-established hemispheric segregation rule of the sign of current helicity in ARs \citep{Seehafer1990, Pevtsov1994, Abramenko1996, Bao1998}. Indeed, as it was argued by \cite{Wheatland2000}, since most of ARs exhibit a non-zero averaged current helicity $\langle B_z J_z \rangle$ the net electric current determined over opposite magnetic polarities must be non-zero and must be of opposite signs.

Since uninterrupted high-spatial-resolution data on vector magnetic fields provided by space-borne instruments (\textit{e.g.}, the \textit{Solar Optical Telescope/Spectropolarimeter} on board \textit{Hinode}, \citealp{Kosugi2007}; the \textit{Helioseismic and Magnetic Imager} on board the \textit{Solar Dynamics Observatory}, SDO/HMI, \citealp{Schou2012}) have become available, the evolution and dynamics of electric current patterns inside ARs were analyzed by many researchers in more detail.
The pattern of electric currents in flare-productive NOAA AR 10930 was studied in a number of works \citep[\textit{e.g.}, ][]{Ravindra2011, Georgoulis2012}. This AR with a strong polarity inversion line (PIL) was associated with emergence of a new magnetic flux and it produced several X- and M-class flares \citep{Kubo2007}. \cite{Ravindra2011} evaluated net electric currents in NOAA AR 10930 separately in positive and negative magnetic polarities. They found that the net electric currents increased in both magnetic polarities simultaneously with the emergence of magnetic flux. Besides, these net currents behaved in exactly the opposite way implying that the current flowed from one polarity to the corona and returned back to the photosphere in the other polarity. The authors attributed variations of net electric currents to the changes of the magnitude of the shear along the PIL where the strongest currents were concentrated.

\cite{Georgoulis2012} used a sophisticated technique to reveal the existence of non-neutralized currents inside ARs. They analyzed the same flare-productive NOAA AR 10930 and flare-quiet AR 10940. AR magnetograms were divided into partitions, with each partition representing a distinct magnetic element of one polarity. The net electric current was calculated in each partition. A partition was assumed to be non-neutralized if the net current in it exceeded an evaluated uncertainty. These authors found that the strongest net current appeared in partitions located along the PIL in NOAA AR 10930. The net currents in NOAA AR 10940 were not as pronounced as in NOAA AR 10930 since the former did not possess a strong PIL. Interestingly, in both ARs magnetic partitions of a given polarity had the same sign of the net electric current.

\cite{Gosain2014} analyzed current patterns in two magnetically-isolated sunspots NOAA 11084 and 11092. High-spatial-resolution data provided by \textit{Hinode}/SOT-SP allowed them to analyze electric current distribution inside sunspot umbra and penumbra. Strong elongated electric current patches of alternating sign were detected along the penumbral fibrils in sunspots. In order to reveal large-scale current patterns associated with the possible global twist, these authors decomposed the electric current into parallel and orthogonal components of the transverse magnetic field. Although the analyzed sunspots exhibited different sense of twist, the current patterns were quite similar. In both sunspots, a strong positive current was found at the central umbral part of the orthogonal component. A thin annulus of a negative electric current outlined the positive umbral current in both cases. The authors suggested that this feature might be associated with return currents.

\cite{Liu2017} studied the relationship between net electric currents, magnetic shear angle, and eruption events in four ARs. They assumed that electric currents of opposite signs within one magnetic polarity represent direct and return currents. Their approach revealed that the ratio of direct to return currents is close to unity in flare-quiet ARs, implying nearly full neutralization. On the other hand, ARs with higher direct/return current ratio exhibited higher level of flare and eruptive activity thus supporting earlier results by \cite{Georgoulis2012} and \cite{Ravindra2011}. These observational findings suggest that flaring activity of an AR may be inversely related to the degree of current neutralization in the AR. This conclusion was further supported in a recent statistical study by \cite{Kontogiannis2017} who compared non-neutralized currents to flare productivity of ARs. Their data covered 336 random days between September 2012 and May 2016 resulting in almost ten thousand of data points. The technique described in \cite{Georgoulis2012} was used to evaluate the net currents and their uncertainties. \cite{Kontogiannis2017} showed that the total net current in an AR is a good predictor of its flare productivity.

The observational detection of return currents is an important milestone for a number of theoretical models that are focused on connection between electric currents and coronal mass ejections (\textit{e.g.}, \citealt{Demoulin2010}, see also Introductions in \citealt{Georgoulis2012} and \citealp{Dalmasse2015}). Simplified theoretical 2.5D models \citep{Dalmasse2015} predict that, with no respect to the exact mechanism of the electric current generation, electric current in a well confined and isolated twisted/sheared magnetic flux tube must be neutralized \citep[\textit{e.g.}, ][]{Melrose1991, Parker1996}. \cite{Melrose1991} further suggested that failure to detect return currents may be explained either by very low magnitudes (below the detection threshold) of these currents that are distributed over a vast area around ARs or by the fact that these currents are very strong and highly concentrated over a small unresolved areas. An alternative explanation is that the return currents may be located below the photosphere (see Figures 3 and 4 in \citealt{Melrose1995}).

Instrumentation limitations could be overcome by sophisticated 3D magnetohydrodynamical (MHD) numerical simulations. Thus, both direct and return currents were found in initially potential magnetic flux tube stressed by photospheric twisting in simulations performed by \cite{Aulanier2005}. Diffuse return currents were observed around each magnetic polarity. The current neutralization problem was addressed in numerical 3D MHD simulations of current-carrying magnetic flux tube emergence carried out by \cite{Torok2014}. In their experiment, an initially current-neutralized magnetic flux buoyantly emerged to the plane stratified atmosphere. A complex redistribution of electric currents observed after the onset of an intense emergence. Predominantly direct currents appeared above the photosphere level resulting in a strong net electric currents in the corona.

Formation of net currents was also scrutinized in a 3D MHD simulations by \cite{Dalmasse2015}, where potential line-tied magnetic fields were stressed by photospheric twisting and shearing motions. These authors argued that buildup of a neutralized electric currents system is an exception.

The above review shows that photospheric electric currents in an AR appear to be non-neutralized. Moreover, the degree of non-neutralization seems to be related to the flaring productivity. In spite of the great importance of this conclusion, from both theoretical and flare-forecast standpoints, the approach does not offer a reliable way to reveal a large-scale, widely distributed electric current system because the requirement of unipolarity might be violated. 
In general, the large significant distributed currents may be spread over a large area with both polarities. How to outline the boundary of such a current system then?

Here we renew an approach, which was proposed earlier in \cite{Abramenko1987} and later elaborated in \cite{Abramenko1991}. We decompose the observed transverse magnetic field vector $\mathbf{B}_{t}$ into two components: parallel to the calculated potential transverse field, and orthogonal to the potential transverse field, $\mathbf{B}_{t \perp}$. The latter, $\mathbf{B}_{t\perp}$ component is generated exclusively by the present electric currents.

In these studies we had found that in some areas of an AR the structure of $\mathbf{B}_{t\perp}$ is well organized and forms a large-scale vortex-like structure (with minor disturbances). Such a structure represents an azimuthal magnetic field associated with a large vertical electric current and a boundary of the vortex encloses the distributed electric current. We emphasize that this approach does not rely on the polarity of magnetic elements covered by the vortex, nevertheless we note that the main spots of an AR are usually located at center of the vortex. With poor-resolution magnetographic data, we reported the magnitude of the distributed current to be of order of 2$\times$10$^{12}$ A \citep{Abramenko1987,Abramenko1991}. In these studies two bipolar ARs located in different hemispheres were analyzed. In each AR, two large vortexes were revealed: one around the leading spot, and another, less pronounced, covered the spots in the following part of the AR. In both ARs, the distributed electric current was directed upward in the leading vortex and downward in the following one. As soon as the (local small-scale) vertical currents were well-balanced in both ARs, a conclusion was made that the distributed upward current of the leading vortex is closed (through the chromosphere and corona) in the following vortex.

In the present paper, we apply this approach to six ARs with different magnetic class. In Section 2, we describe data selection and reduction, a procedure of calculation of local (resolution-scale) and large distributed currents is discussed in Section 3, analysis of time variations of currents and other AR parameters is presented in Section 4, and our conclusions are listed in the last section.

\section{Data Selection and Reduction}
\label{S-methods}

For our study we selected six ARs listed in Table~\ref{table1}. The guideline for selection was as follows. First, the set must represent both flare-quiet and flare-productive ARs. The top three ARs listed in Table~\ref{table1} are low-flaring groups. The other three exhibited enhanced flaring activity. The ARs in Table~\ref{table1} are ordered by the increasing flare index, FI (4\textsuperscript{th} column in Table~\ref{table1}), which represents flare productivity of an AR \citep{Abramenko2005} and equals 1 (100) for an AR, that produced one C1.0 (X1.0) flare per day. Second, flaring ARs should represent the essential magnetic structures. Thus, AR NOAA 12158 is an anti-Hale group (with wrong leading polarity), 12371 is a bipolar and 12192 is a multipolar group.

The main data source used in this work was SDO/HMI vector magnetic field measurements provided by the Joint Science Operation Center (JSOC, \verb|http://jsoc.stanford.edu/|). SDO/HMI is a 4096$\times$4096 pixel full-disk filtergraph that routinely performs measurements of full Stokes vector in Fe~I 6173 \AA\ spectral line \citep{Schou2012, Liu2012}. The Stokes vector measurements are used to derive full-disk vector magnetograms, Dopplergrams, and other quantities with a cadence of 720 s. The spatial resolution of the instrument is 1 arcsec with the pixel size of 0.5$\times$0.5 arcsec\textsuperscript{2}. A special algorithm is used \citep{Turmon2010} to automatically identify and crop ARs patch from the full-disk magnetograms. The patches of ARs are provided as Space-Weather Active Region Patches \citep[SHARPs][]{Bobra2014, Hoeksema2014} that include maps of magnetic field strength, $B_f$, inclination, $B_i$, and azimuth, $B_a$. 

To minimize the influence of the projection effect we tracked ARs as long as they were located within $\pm$35 degrees of the central meridian (corresponding to a four-day time interval, which are shown in the 3\textsuperscript{rd} column of Table~\ref{table1}). Assigning the $z$-axis to coincide with the line-of-sight (LOS) component of the magnetic field, we calculated all components of the magnetic field vector as

\begin{equation}
B_x = B_f\sin(B_i)\sin(B_a); \\
B_y = -B_f\sin(B_i)\cos(B_a); \\
B_z = B_f\cos(B_i).
\label{eq3}
\end{equation}

The averaged total unsigned flux, $\langle \Phi \rangle$, that an AR exhibited during the analyzed period is shown in the 5\textsuperscript{th} column of Table~\ref{table1}.

Flare activity of the ARs was evaluated using the 1--8 \AA\ X-ray flux measurements acquired by the \textit{Geostationary Operational Environmental Satellite-15} (GOES-15, the data are available at \verb|https://satdat.ngdc.noaa.gov/sem/goes/data/full/|). We also used images of the Sun acquired at the 1600 \AA\ spectral line by the \textit{Atmospheric Imaging Assembly} on board SDO \citep[SDO/AIA,][]{Lemen2012}, to analyze the structure and dynamics of the associated flares.

\begin{table}
	\centering
	\caption{Parameters of ARs under study}
	\label{table1}
	\begin{tabular}{cccccccccc} 
		\hline
		NOAA   & Lat.   & Obs.     & FI & $\langle \Phi \rangle $, & $\langle I_{tot} \rangle$,& $\langle I_{net} \rangle$, & $\langle I_{distr}\rangle$,  &$\langle \rho_{j_{z}} \rangle$, & $\langle \rho_{B_{z}} \rangle$ \\
		number & deg. & interval &    & $10^{22}$ Mx              & $10^{15}$ A                 & $10^{12}$ A                    & $10^{12}$ A & \% & \%\\
		\hline

		12674	& N14 &	 2017 Sep 03--06 & 0.76 & 2.47   & 3.74 & -1.27	&  5.98 &  -0.034 &  -9.270\\
		12494	& S12 &	 2016 Feb 05--07 & 1.02 & 0.73   & 1.13 & 0.37  & 8.23  &  0.033  &  -14.757\\
		12381	& N14 &	 2015 Jul 07--10 & 5.43 & 1.38   & 2.37 & -1.78 &  2.91 & -0.075 &  0.039 \\
		\hline
		12158	& N15 &	 2014 Sep 09--12 & 13.30 & 1.45  & 2.48 & 0.54 & -12.53 &  0.022 &  3.970 \\
		12371	& N13 &	 2015 Jun 20--23 & 20.13 & 2.97  & 3.36 & 3.26  & 23.60 & 0.097  &  2.125 \\
		12192	& S14 &	 2014 Oct 22--25 & 123.44 & 9.53 & 10.51& 6.52  & 58.14 &-0.062 &  -2.074\\

		\hline
\end{tabular}
\end{table}

\section{Local and Distributed Electric Currents}

One way to derive the magnitude of electric current density $j_z$ within an AR is to use the differential form of Ampere's law to compute a local electric current map, \textit{i.e.}:

\begin{equation}
j_z = \frac{1}{\mu_0}\left( \frac{\partial B_y}{\partial x} - \frac{\partial B_x}{\partial y} \right),
\label{eq1}
\end{equation}
where $\mu_0$ is the magnetic constant, and $B_y$ and $B_x$ are components of the transverse magnetic field $\mathbf{B}_t$. It is worth noting that due to shortcomings of magnetographic instrumentation this approach may not be well suitable for deriving electric current density \citep{Parker1996}. Moreover, \cite{Fleishman2018} argued that the use of partial derivatives may lead to enhancement of errors of calculations.

In this work we used the integral form of the Ampere's law to compute the electric current density at each pixel of a magnetogram:

\begin{equation}
j_z = \frac{1}{\mu_0 s} \oint_L{\mathbf{B}_t d\mathbf{r}},
\label{eq2}
\end{equation}
where the integration is performed over a small closed contour $L$ enclosing an area $s$ where the vertical electric current density is to be calculated. Our previous study \citep{Fursyak2018} showed that the contour size of 5$\times$5 pixels is
the best compromise between the noise level and loss of information due to smoothing by the integration. Thus, larger contours produce less intense and wide current structures whereas a smaller contour produces no visible improvement compared to the outcome of the differential formula. When using the 5$\times$5 pixel contour, the width of peaks remains the same and the noise level is much lower than that of the differential method. 
That is because to calculate the electric current density at the central pixel, the integral method uses 16 nodal points (for $L$=5$\times$5 pixels) versus only 9 points are used in the differential method. Moreover, integration is performed using the Simpson formula to enhance the calculation accuracy. The large-scale distributed currents were then derived by summing the current density inside the area of interest.

Note, that the Ampere's law was used for investigation of neutralized currents as means of computing large-scale currents, distributed over an area covering magnetic field of one polarity \citep{Wilkinson1992, Georgoulis2012, Kontogiannis2017}. An integration of the transverse magnetic field along a vast contour was applied. To derive the current densities in nodal points of a magnetogram, this method was applied for the first time by \cite{Abramenko1987} and \cite{Abramenko1988}. Here we continue this way to utilize the Ampere's law. 

\begin{figure}    
	\includegraphics[width=1.\textwidth,clip=]{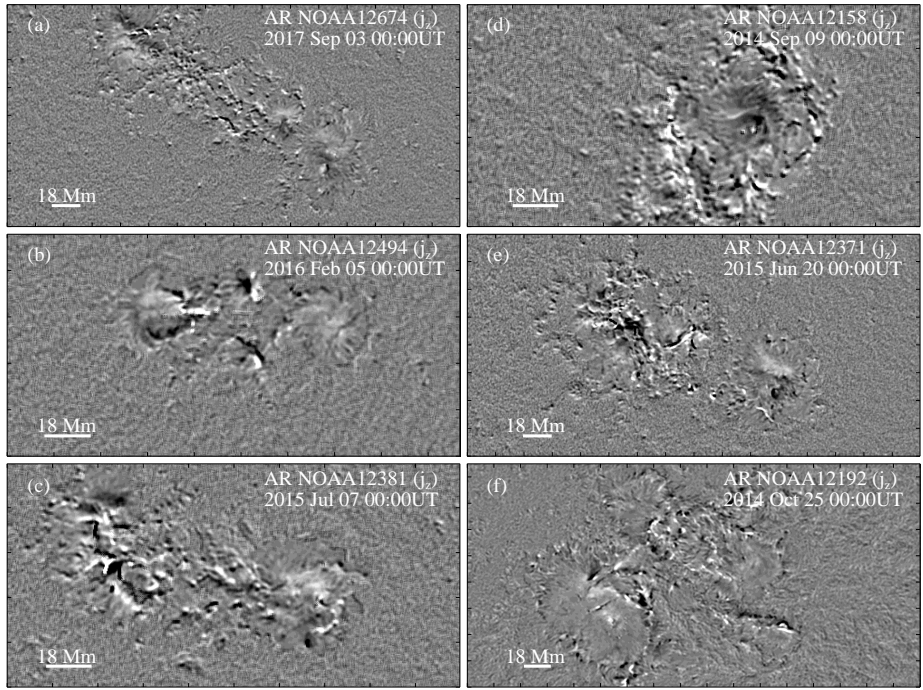}
	\caption{ Typical maps of electric current density in NOAA ARs 12674 (a), 12494 (b), 12381 (c), 12158 (d), 12371 (e), and 12192 (f) derived from the transversal magnetic field data by equation \ref{eq2}. The maps are scaled from -0.02 A m$^{-2}$ (black) to 0.02 A m$^{-2}$ (white).
	}
	\label{fig1}
\end{figure}

Typical distributions of local vertical electric currents in ARs are shown in Figure~\ref{fig1}. For each magnetogram acquired during the analyzed period we calculated the unsigned total vertical current, $I_{tot}$, as a sum of absolute values of the current density multiplied by the pixel area. The averaged over time $I_{tot}$ are presented in the 6\textsuperscript{th} column of the Table~\ref{table1}. We can see that the magnitudes are nearly similar (except for the strongest AR 12192), suggesting that the flare-quiet and flare-productive ARs do not significantly differ.

We also calculated the imbalance of the local currents and the magnetic flux (their time-averaged values are listed in the last two columns of Table~\ref{table1}). We used the commonly accepted formula for the imbalance \citep{Abramenko1996}:

\begin{equation}
\rho_{j_{z}} = \frac{\sum_{S+} |j_z(i,j)| - \sum_{S-} |j_z(i,j)|}{\sum_{S+} |j_z(i,j)| + \sum_{S-} |j_z(i,j)|} \times 100 \%,
\label{eq4}
\end{equation} 
where $j_z(i,j)$ is electric current density at pixel $(i, j)$, $S+$ ($S-$) denotes a set of pixels with $j_z(i,j)>0$ ($j_z(i,j)<0$).

 Using the same approach we also calculated the imbalance of the vertical component of the magnetic field, $\rho_{B_{z}}$ (last column in Table~\ref{table1}). The numerator in Equation~(\ref{eq4}) gives us the net current over a magnetogram, and its time-averaged magnitude $<I_{net}>$ is also presented in the 7\textsuperscript{th} column of Table~\ref{table1}.

Comparison of the last two columns of Table~\ref{table1} demonstrates that, for all ARs, the current imbalance is very low (it does not exceed 0.1\% percent), whereas the flux imbalance can be quite strong. This implies that the vertical electric currents are closing within an AR, whereas significant fraction of the magnetic flux of an AR may close elsewhere outside the AR. This is a common situation especially during the solar maximum since UV images often show large-scale loops connecting various ARs or an AR and a quiet Sun area. Moreover, this also means that electric currents do not always flow along the magnetic field lines, \textit{i.e.}, the photospheric magnetic field is not a force-free field and/or a substantial part of the magnetic flux is nearly potential and leaves the AR.

Structured local electric currents and elongated current ribbons of both signs could be seen in the current distribution maps in Figure~\ref{fig1}. If the surface-distributed current does exist, it is not readily visible in these maps. To reveal it we used a method introduced and tested in \cite{Abramenko1987}. A distributed vertical electric current can manifest itself as a regular deviation from potentiality, \textit{i.e.}, as an organized vortex-like azimuthal magnetic field. Therefore, one may detect distributed electric currents by analyzing the deviation of the observed magnetic field lines from the corresponding potential configuration.

\begin{figure}    
	\includegraphics[width=1.\textwidth,clip=]{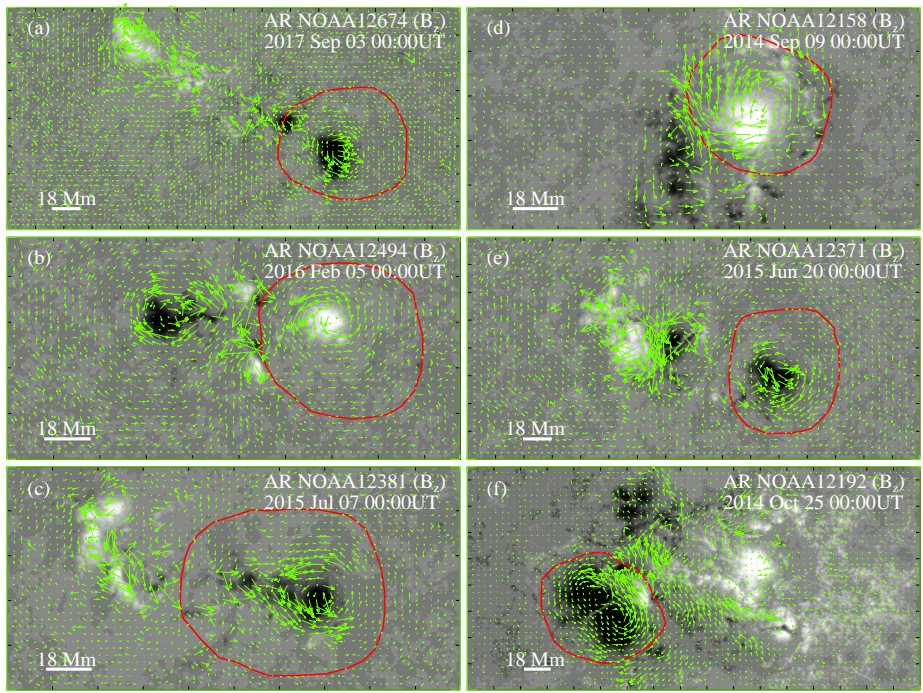}
	\caption{Line-of-sight magnetograms and the vector of the non-potential component of the transversal magnetic field $\mathbf{B}_{t\perp}$ (green arrows) for NOAA ARs 12674 (a), 12494 (b), 12381 (c), 12158 (d), 12371 (e), and 12192 (f). Red thick curves show the chosen contours $C$ used to calculate the magnitude of distributed electric current (see text).
	}
	\label{fig2}
\end{figure}

We thus performed the following procedure. For each vector magnetogram, we calculated a potential magnetic field based on the observed $B_z$ component by IDL CFF1N code \citep{Sakurai1982}. At each pixel of the magnetogram, the observed transverse magnetic field vector ($\mathbf{B}_t$) was decomposed into two components: a component parallel to the transverse potential magnetic field and a component, $\mathbf{B}_{t\perp}$, that is perpendicular to the transverse potential magnetic field. The latter is generated by vertical electric currents. We will refer to $\mathbf{B}_{t\perp}$ as a non-potential component of the transverse magnetic field (Figure~\ref{fig2}, green arrows).

 Figure~\ref{fig2} shows that a vortex-like structure can be detected in each AR. Some of them are very well pronounced (especially those around the main spot), whereas others, spread over the following part of an AR (except NOAA AR 12192), demonstrate only a hint of a vortex-like structure with an opposite sense of twist. The most complex AR 12192 (Figure~\ref{fig2}f) displays at least three vortex-like structures with the strongest one associated with the strongest negative polarity spot. Note, that these vortexes are not exclusively connected to any particular magnetic polarity, but may cover an extended area encompassing both polarities, see Figure~\ref{fig2}d,~f.
	
We speculate that an observed vortex is associated with a large-scale distributed vertical electric current, which must close, in a loop-like manner, through the chromosphere and the corona back to the photosphere, because the vertical electric currents are very well balanced (see Table~\ref{table1}, column 9).

We thus possibly deal with a global electric current system of an AR, and to evaluate the magnitude of this current, we need to integrate current densities inside the area of the most coherent and strong vortex. In case of bipolar ARs, the current seems to be closed predominantly over the vast following part: the hints of opposite vortexes observed there support the suggestion. In a case of very complex multipolar magnetic configurations (such as AR NOAA 12192), several of such global current systems might coexist in one AR. 

To this end, to calculate the magnitude of the distributed electric current, we have to focus on a best-pronounced and strongest vortex of $\mathbf{B}_{t\perp}$. 
We calculate the magnitude of distributed electric current as follows:

\begin{equation}
I_{distr} = \int_S{j_z ds},
\label{eq5}
\end{equation} 
where $S$ is an area enclosed by a contour $C$ (note that the electric current densities in Equation~(\ref{eq5}) are taken with their sign). This contour was manually defined such that the $\mathbf{B}_{t\perp}$ arrows are oppositely directed on both sides of the contour outline. The contour was defined on the first magnetogram, the shape of the contour was kept the same during the observational interval and the location of the contour was fixed relative to the center of gravity of the sunspot. The center of gravity was measured in magnetograms, the contour of the main sunspot was determined as a level of $\pm$ 1000 Gauss in the vertical magnetic field component $B_z$. 
The summing (\ref{eq5}) was performed over all pixels located inside the contour $C$, 
 regardless of the polarity of $B_z$. Note that the vertical magnetic field component does  not pertain directly to calculations of $I_{distr}$, only indirectly through the determination of the contour C (as a boundary condition for the potential field calculations). At the same time, one can notice (see Figure~\ref{fig2}) that small places of opposite (relative to the main spot) polarity inside the contour are co-spatial with local disturbances of the coherent $\mathbf{B}_{t\perp}$-vortex. We consider appearance/disappearance of such disturbances as a signature of evolutional changes in the distributed current.

 The sign of $I_{distr}$ is positive when the resulting current is directed toward the observer and negative in the opposite case. The time-averaged $I_{distr}$ for each AR are presented in  the 8\textsuperscript{th} column of Table~\ref{table1}.
Data in this table show that for all ARs the net current over a magnetogram, $I_{net}$, is weaker than the net current inside a contour,  $I_{distr}$. This implies that the outlining a specific contour to detect the global structure with prevailing current really makes sense.

The direction of the distributed electric current can also be derived from the direction of the vortex structure of the non-potential transverse magnetic field: a distributed electric current directed toward the observer is associated with predominantly counterclockwise direction of $\mathbf{B}_{t\perp}$.

We can estimate the sign of magnetic twist, $\alpha$, from the well-known relationship: $j_z = \alpha B_z$. In general, the direction of twist is related to magnetic helicity of a magnetic flux system. A number of various mechanisms is presumably responsible for helicity generation within a magnetic flux tube (Coriolis effect, differential rotation, turbulent plasma motions to mention a few). Most of these mechanisms are antisymmetric with respect to the solar equator, therefore the twist of ARs in different hemispheres should have opposite signs. This inference, known as the hemispheric helicity sign rule, was observationally confirmed in a number of works (\textit{e.g.}, \citealp{Seehafer1990, Pevtsov1994, Abramenko1996}, see also a review by \citealp{Pevtsov2014}). ARs obeying the hemispheric rule should have counterclockwise (clockwise) direction of the $\mathbf{B}_{t\perp}$ arrows in the leading (following) parts regardless of the hemisphere where they are located.

For ARs 12381 and 12674, located in the northern hemisphere, the $B_z$ of the leading spot is negative, therefore, magnetic twist, $\alpha$, is also negative in accordance with the hemispherical helicity rule.
The upward (positive) distributed current in the leading part is in accordance with the relationship $j_z = \alpha B_z$.

For NOAA AR 12494 located in the southern hemisphere, we observed a positive polarity leading sunspot and a positive effective magnetic twist, as is expected for an AR in the southern hemisphere according to the hemispherical helicity rule. Accordingly, we observed the upward distributed current in the leading part of the AR.

The situation is more diverse for flaring ARs. Thus, in the case of NOAA AR 12371, which is a bipolar AR and obeys both the Hale-polarity law and the hemispheric helicity rule, the entire picture is similar to that for the flare-quiet ARs with one exception: the magnitude of the distributed current is much higher, see the 8\textsuperscript{th} column in Table~\ref{table1}.
	
NOAA AR 12158 was located in the northern hemisphere, however, contrary to Hale polarity law, it had a positive polarity leading spot. The distributed current over this positive polarity leading part turned out to be negative, which resulted in the negative helicity $\alpha$, which means that the AR obeyed the hemispheric helicity rule. Indeed, the corresponding EUV images indicated the overall counter-clockwise twist of coronal loops around the leading spot.

For NOAA AR 12192, we found an upward distributed electric current around the main following spot, which implies that this AR does not obey the hemispheric helicity rule. This peculiar and the largest AR of solar cycle 24 \citep{Sheeley2015} is discussed in detail in the next section.

The most interesting inference that follows from Table~\ref{table1} is that the magnitude of the distributed current found in the flaring ARs significantly exceeds that of the flare-quiet ARs. This fact motivated us to further explore time variations of the distributed current and compare them to the activity time lines of these ARs.

\section{Temporal Variations of the Distributed Current in ARs of Different Flare Activity}

\begin{figure}    
	\includegraphics[width=1.\textwidth,clip=]{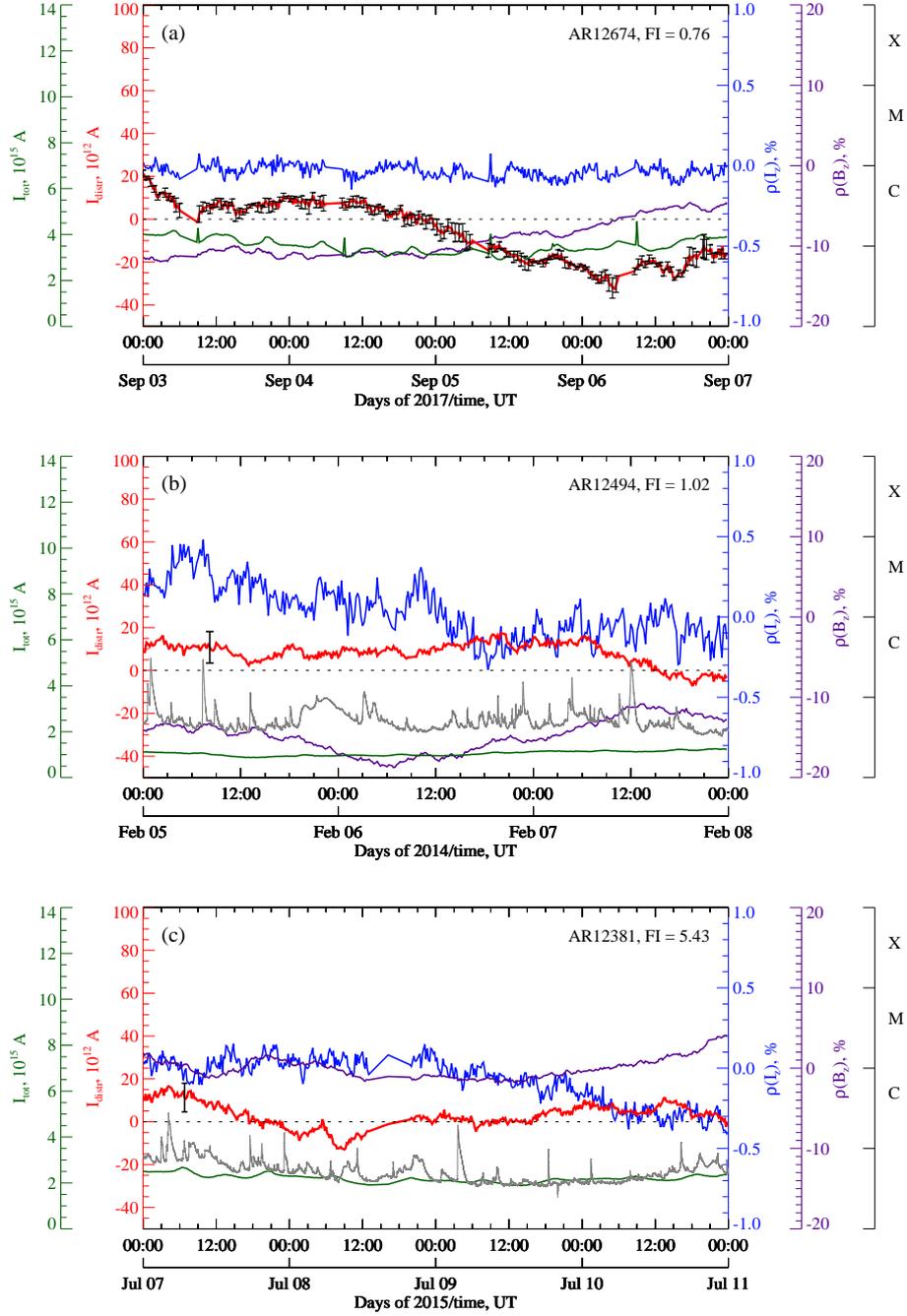}
	\caption{
		Time variations of the distributed electric current magnitude (red), total electric current (green), electric current imbalance (blue), and magnetic flux imbalance (violet) for a sample of flare-quiet ARs. GOES-15 X-ray flux is shown in grey. On the top panel the GOES flux is omitted because a strong domination of the neighbor NOAA AR 12673. Error bars in calculations of $I_{distr}$ for each third magnetogram are shown for AR 12674, the highest error is marked as a solid bar. For the rest of ARs, only the highest error bar is shown. 
	}
	\label{fig3}
\end{figure}

\begin{figure}    
	\includegraphics[width=1.\textwidth,clip=]{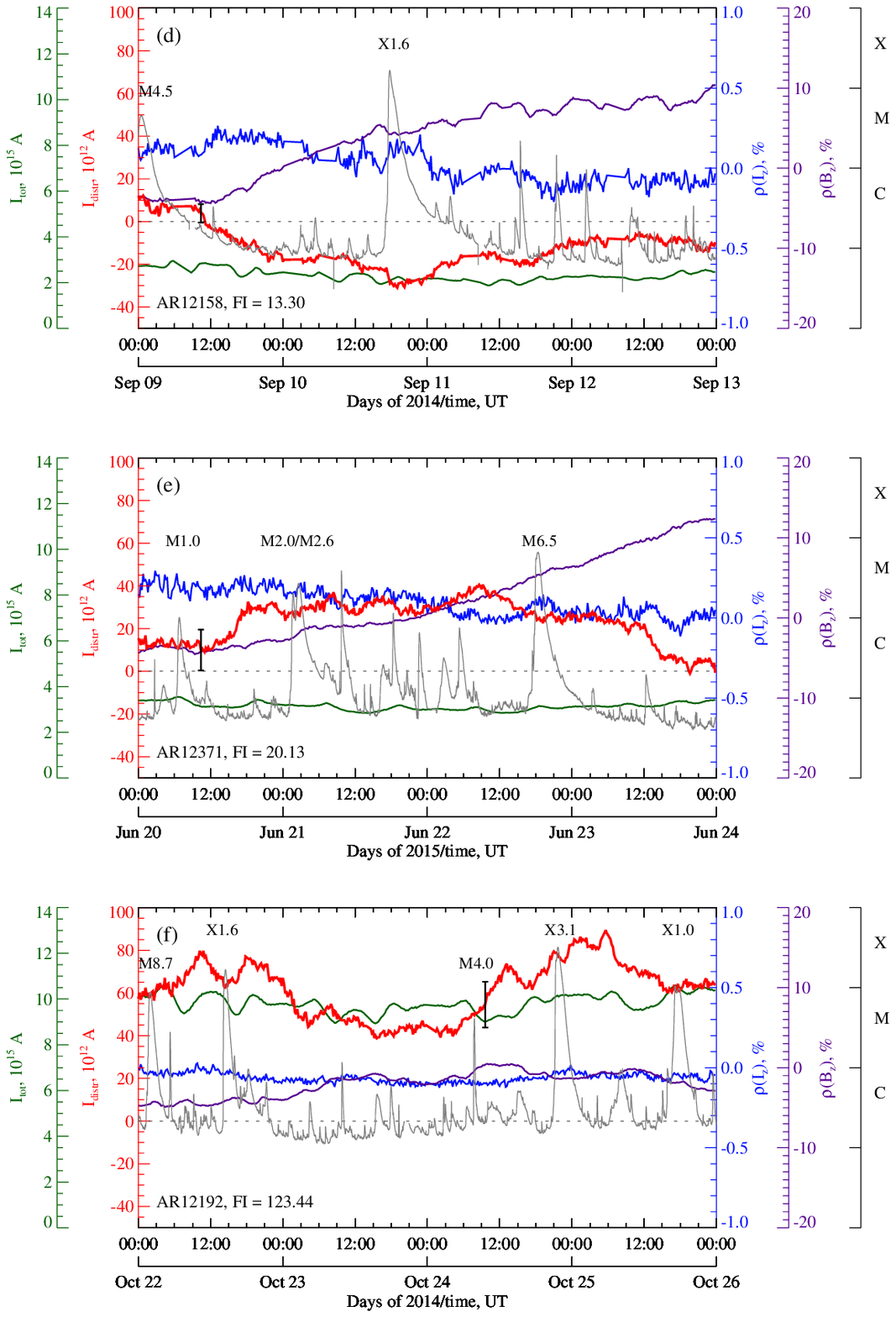}
	\caption{
		The same as in Figure~\ref{fig3} for a sample of flare-productive ARs. GOES-15 X-ray flux is shown in grey. The strongest flares occurred in a given AR are marked. Other notations are the same as in Figure \ref{fig3}.
	}
	\label{fig4}
\end{figure}

We will study time variations of the distributed electric current, $I_{dist}$, inside the contours marked in Figure \ref{fig2}. The time variations of the distributed electric current in flare-quiet ARs, along with other AR parameters are shown in Figure~\ref{fig3}, while those for flaring ARs are presented in Figure \ref{fig4}.
We found that the largest errors in calculations of $I_{distr}$ come from possible variations in the manually-derived contour C. For each AR, several contours were applied, and the largest deviation in  $I_{distr}$ was adopted as an error bar. The error bars for each third magnetogram of NOAA AR 12674 are shown in the top panel of Figure \ref{fig3}, along with the highest (during a 4-day interval) error bar, which is  marked as a solid bar. For the rest of ARs, only the highest error bar is shown in each panel of Figures \ref{fig3}, \ref{fig4}.

We found that in all ARs the dominant orientation of the $\mathbf{B}_{t\perp}$ vortex 
(counterclockwise or clockwise)
was largely preserved during the observed period. However, the magnitude of the distributed current was changing with time. Thus, flare-quiet ARs (see Figure \ref{fig3}, red curves), show a rather low magnitude of $I_{distr}$ (in a range of $\pm 20 \times 10^{12}$ A), and the sign of $I_{distr}$ can change as well. This behavior may be explained by intrusion of new strong small-scale local electric currents into the vortex area driven by sudden appearance/disappearance of magnetic features.

Flaring ARs show much higher level of $I_{distr}$ (up to $80\times10^{12}$ A), so that the possible disturbances by small-scale current features do not affect the sign of $I_{distr}$ and time variations of the $I_{distr}$ magnitude are very gradual. For two ARs (12158 and 12192), we observed slight enhancements of the distributed current, which were co-temporal with periods of enhanced flaring.

We did not find any one-to-one correspondence in time variations between the total unsigned current $I_{tot}$ and the distributed current $I_{distr}$. This is not surprising because $I_{distr}$ constitutes only a tiny part (of about one thousandth) of $I_{tot}$ (see Table~\ref{table1} and Figures \ref{fig3}, \ref{fig4}). Nevertheless, energy stored in the distributed current system of about $(5-80) \times$ $10^{12}$ A is about $10^{32}-10^{33}$ erg, which is comparable with energy released in any solar flare.
The energy magnitude was estimated following \cite{Abramenko1987}: an azimuthal field $B_{\phi}(r)$ of the electric current $I$ uniformly distributed over the cross-section of a cylinder of radius $a$ can be presented as $B_{\phi}(r)=(2I/ca^2)/r$. Here $r$ is a distance from the cylinder center  ($0<r<a$) and $c$ is the speed of light. An integration of $B_{\phi}^2(r)/(8\pi)$ over the volume occupied by the current loop of typical length $10^{10}$ cm and typical radius $5\times10^9$ cm gives us the estimation of distributed electric current energy.

The strongest (in sense of the total magnetic flux and the flaring index) NOAA AR 12192 deserves more attention as this AR seemed to host two large-scale distributed current systems.

\begin{figure}    
	\includegraphics[width=1.\textwidth,clip=]{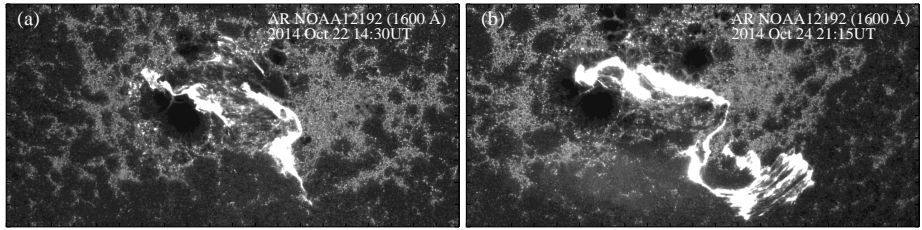}
	\caption{
	UV 1600 \AA\ SDO/AIA images of NOAA AR 12192 during flare events X1.6 on 2014 October 22 (a) and X3.1 on 2016 October 24 (b). A flare ribbon in the south-west peripheral area of the AR is clearly seen during the X3.1 event.
	}
	\label{fig5}
\end{figure}

We thus speculate that the first distributed electric current system appeared to connect the leading positive polarity sunspot with several small sunspots of negative polarity located in the northern part of the AR. The direction of that distributed electric current obeyed the hemispheric helicity rule.
The second and prevailing distributed electric current system could have connected the mature following negative polarity sunspot with a vast positive polarity region located to the south-west
(the region is inside the field-of-view, see Figure~\ref{fig2}f, and, therefore, it was included in our calculation of the electric current imbalance).
This assumption is supported by data in Figure~\ref{fig5} showing flare ribbons connecting the two discussed regions. 
Flare ribbons are usually associated with eruption of a large scale twisted current carrying structure as inferred from data analysis by \cite{Sun2015}, as well as from numerical simulations by \citet{Jiang2016}.

We analyzed time variations of the second (prevailing) distributed electric current system in NOAA AR 12192 associated with the mature following sunspot (Figure~\ref{fig2}f).
 Our choice was dictated by two facts. First, the transverse field vortex was more pronounced (in sense of stronger transversal field) in this part of the AR allowing us to more accurately define the boundary contour $C$. Second reason is that at least one flare brightening associated with strong flares in this AR was located in the vicinity of the following sunspot. The current time profile shows that the magnitude of the distributed current in NOAA AR 12192 is noticeably higher than those found in other flaring ARs under study (Figure~\ref{fig4}).
Note, that NOAA AR 12192 also displayed the highest flare index (see Table~\ref{table1}, column 4). Variations of the distributed electric current are also more pronounced than those for other ARs.
Thus we clearly observe that periods of enhanced flaring (M- and X-class flares) are nearly co-temporal with the high level of distributed electric current.

\begin{figure}    
	\includegraphics[width=1.\textwidth,clip=]{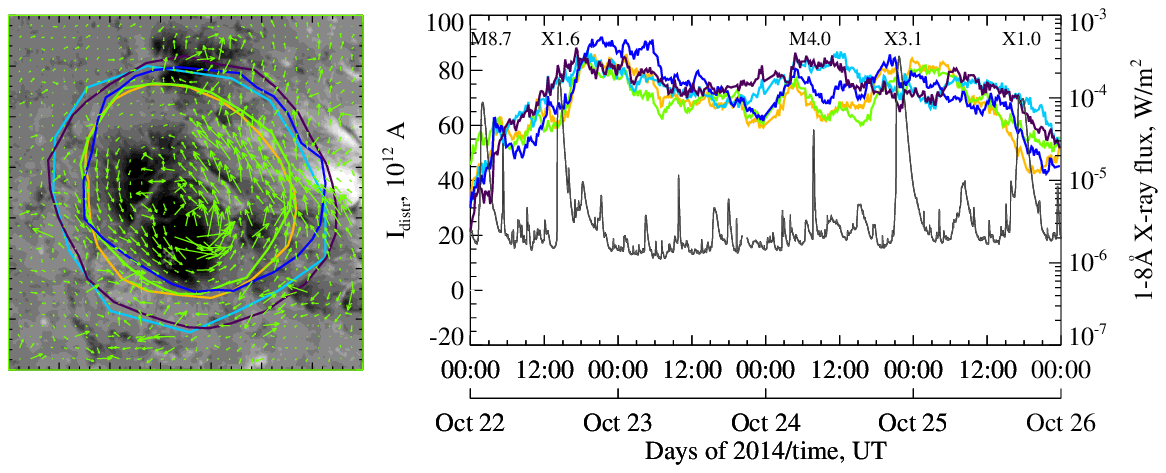}
	\caption{
		Left -- LOS magnetogram and the vector of the non-potential component of the transversal magnetic field $\mathbf{B}_{t\perp}$ (green arrows) of the following polarity of NOAA AR 12912. Colored curves show different manually chosen contours $C$ to calculate the distributed electric current magnitude. Right -- Time variations of the distributed electric current magnitude are shown by colored thick curves. The color of each curve denotes the contour that was used to calculate the magnitude of distributed electric current by equation~\ref{eq5}. GOES-15 X-ray flux is shown by black curve. The strongest flares occurred in NOAA AR 12192 are marked.
	}
	\label{fig6}
\end{figure}

It should be noted that the magnitude of distributed electric current as well as its temporal behavior only slightly depended on the shape of the boundary contour $C$. In Figure~\ref{fig6} we show a set of different contours that were used to calculate $I_{distr}$ and the corresponding temporal variations of $I_{distr}$ are shown in the right panel of Figure~\ref{fig6}. One can see that all time curves exhibit similar behavior and amplitudes.

\section{Concluding Remarks and Discussion}

Using SDO/HMI magnetic field data for six ARs, we studied large scale organization of vertical electric currents. Our findings are as follows.
	
\begin{enumerate}
	
\item  In all cases the imbalance $\rho_{j_{z}}$ of local vertical currents over the entire magnetogram was found to be very low (from 0.02 to 0.1\%) and it remained persistently low during the entire studied time interval (4 days). At the same time, imbalance of the vertical magnetic field was much higher (up to 14\%). This finding implies that, first, vertical electric currents are closed inside the AR and, second, the magnetic field and electric currents do not always follow each other, so that the photospheric magnetic field either is not a force-free field, or a part of the magnetic flux is potential and leaves an AR.

The highly fragmented structure of the local currents allowed us to suggest that the bulk of the current is closed within low lying loops. This suggestion is supported by NLFFF modeling of magnetic field and currents above an AR 10930 \citep[][see Fig. 3 in this paper]{Schrijver2008}.
Numerical simulations \citep[][to mention a few]{Georgoulis2012, Janvier2014, Dalmasse2015} showed that places with significant shear and strong magnetic polarity inversion line are particularly associated with strong local electric currents, that seem to be closed in short low-lying magnetic loops.
Using data-driven numerical magnetohydrodynamic (MHD) modeling, \cite{Jiang2016} found that in the AR of our interest NOAA AR 12192, the essential electric currents are concentrated below $\approx$20 Mm \citep[see Fig.~2 in][]{Jiang2016}. 
A logical consequence of this result is that the electric current density decreases with height, so that high loops (including those closing outside the AR and carrying out magnetic flux) are nearly potential. This is supported by \cite{Abramenko1997}, and by \cite{Schrijver2016} studies, who compared potential field configuration with observed EUV loops for an extended sample of ARs.

Final and the most important inference of this study is that an electric current system of any scale is closing within an AR.

There are only few studies of the electric currents imbalance over an entire AR. Thus, \cite{Abramenko1996} analyzed 40 ARs using \textit{ Huairou Solar Observing Station} (HSOS) data and found that the imbalance of the vertical electric current is low (maximal imbalance of 3.5\% and it was lower than 1\% for 26 ARs). \cite{Schrijver2008} reported the electric current imbalance of NOAA AR 10930 to be 0.7\%, while \cite{Georgoulis2012} reported 6.3\% imbalance for NOAA AR 10940, which is four times lower than the magnetic flux imbalance.

Thus, the present study further supports the earlier finding that the vertical electric currents in ARs are very well balanced.

\item 

Structures of electric currents of different spatial scales co-exist within an AR. Along with local currents, we found an active region-scale distributed current $I_{distr}$ associated with large-scale coherent vortexes of the non-potential component $\mathbf{B}_{t\perp}$ of the observed transverse magnetic field. The leading part of an AR is usually occupied by the upwardly directed distributed current, which closes down to the photosphere through the chromosphere and the corona over a vast area of the following part of the AR.
	
Note that the $\mathbf{B}_{t\perp}$-vortex (over the leading or following part of an AR) covers a large area that encompasses magnetic elements of both polarities. This is essentially different from the neutralized current investigations.

\item The magnitude of the distributed electric current in an AR differs depending on the level of flare activity. Thus, low-flaring ARs exhibit small variations of the magnitude of the current system in the range of $\pm20\times 10^{12}$ A while ARs with high level of flaring show significant variations of the distributed electric current in the range of $30-95 \times 10^{12}$ A.
 Qualitatively, this inference is in a good agreement with the earlier findings on the non-neutralized electric currents  \citep{Kontogiannis2017}, however, there should be some caution in the interpretation. Non-neutralized electric currents are associated with the Lorentz force in the vicinity of polarity inversion lines \citep{Georgoulis2012} and thus emphasize local concentrations of electric currents closing in relatively low loops. In the present study, global electric current systems are investigated, which are, in principle, not associated with the polarity inversion lines and seem to penetrate in the corona. An association between these current systems is a subject of future investigations.

 When compared to the total unsigned electric current in an AR, the magnitude of the distributed current is rather small, less than one percent of the total current. However, energy stored in the distributed electric current of $(5-80)\times 10^{12}$ A found in a typical AR is about $10^{32}-10^{33}$ erg, which is comparable with energy of any solar flare.

\item  We also found that periods of high level of the distributed electric currents are nearly co-temporal with enhanced flaring in an AR.
	
	The magnitude of the distributed currents varies rather gradually with a characteristic time-scale of several days, which agrees with \cite{Melrose1991} and \cite{Wheatland2000} who argued that an AR-scale current system is rooted deeply in the convection zone, and ``the long inductive time associated with such an extended current system precludes changes on the short time scale of a flare, and so currents will be conserved during a flare’’ \citep{Wheatland2000}.
	In this sense, our inferences are in agreement with the so-called [$\mathbf{E}$-$\mathbf{J}$] paradigm, which considers magnetic loops as deeply rooted electric circuits \citep{Melrose1995}, on the contrary to the [$\mathbf{B}$-$\mathbf{u}$] paradigm (Parker 1996) based on the magnetohydrodynamical description of the magnetic field $\mathbf{B}$ and velocity field $\mathbf{u}$ \citep[see][]{Georgoulis2018}. This long-standing dichotomy between the interpretations, where the non-neutralization of electric currents plays a key role, seems to be far from a solution, however, recent results reported by \citet{Georgoulis2012} and \citet{Georgoulis2018} provided solid criteria and physical reasons for observed non-neutralization of electric currents.


\end{enumerate}	
	
	Results presented here are based on data from six ARs. Although the most representative types of ARs were analyzed further investigation involving larger statistics is necessary in order to understand the role of the large-scale global current systems in AR stability and flaring.

\begin{acks}

 Authors thank Referee for interesting comments, criticism, and useful suggestions.
SDO is a mission for NASA's Living With a Star (LWS) program. The SDO/HMI data were provided by the Joint Science Operation Center (JSOC). The study was supported by the Russian Science Foundation (Project 18-12-00131).

\end{acks}



\end{article} 

\end{document}